\documentclass[copyright,creativecommons]{eptcs}
\providecommand{\event}{GandALF 2015}
\usepackage[T1]{fontenc}
\usepackage{nico}
\usepackage{avg}
\usepackage{microtype}

\title{Average-energy games\thanks{Work partially supported by
    European project CASSTING (FP7-ICT-601148) and ERC project EQualIS
    (StG-308087).}}
\author{Patricia~Bouyer \qquad Nicolas Markey \qquad Mickael
  Randour\institute{LSV -- CNRS \& ENS Cachan -- France} \and
Kim~G.~Larsen \qquad Simon Laursen\institute{Aalborg University -- Denmark}}
\def\titlerunning{Average-energy games}
\def\authorrunning{P.~Bouyer, N.~Markey, M.~Randour, K.G.~Larsen, S.~Laursen}

\begin{document}

\maketitle

\begin{abstract}
Two-player quantitative zero-sum games provide a natural framework to
synthesize controllers with performance guarantees for reactive systems within
an uncontrollable environment. Classical settings include mean-payoff games,
where the objective is to optimize the long-run average gain per action, and
energy games, where the system has to avoid running out of energy.

We study \textit{average-energy} games, where the goal is to optimize the
long-run average of the accumulated energy. We show that this objective arises
naturally in several applications, and that it yields interesting connections with
previous concepts in the literature. We~prove that deciding the winner in such
games is in \NP $\cap$ \coNP and at least as hard as solving mean-payoff
games, and we establish that memoryless strategies suffice to~win. We~also
consider the case where the system has to minimize the average-energy
\textit{while} maintaining the accumulated energy within predefined bounds at
all times: this~corresponds to operating with a finite-capacity storage for
energy. We give results for one-player and two-player games, and establish
complexity bounds and memory requirements.

\end{abstract}

\section{Introduction}
\label{sec:intro}
%

\paragraph{Quantitative games.} Game-theoretic formulations are a standard
tool for the synthesis of provably-correct controllers for reactive
systems~\cite{GTW02}. We consider two-player (system vs. environment)
turn-based games played on finite graphs. Vertices of the graph are called
\textit{states} and partitioned into states of player~1 and states of
player~2. The game is played by moving a pebble from state to state, along
\textit{edges} in the graph, and starting from a given initial state. Whenever
the pebble is on a state belonging to player~$i$, player~$i$ decides where to
move the pebble next, according to his \textit{strategy}. The infinite path
followed by the pebble is called a \textit{play}: it~represents one possible
behavior of the system. A~\textit{winning objective} encodes acceptable
behaviors of the system and can be seen as a set of winning plays. The goal of
player~1 is to ensure that the outcome of the game will be a winning play,
whatever the strategy played by his adversary.

To reason about resource constraints and the performance of strategies, \textit{quantitative games} have been considered in the literature. See for example~\cite{emsoft2003-CAHS,BCHJ09,Ran13}, or~\cite{Ran14} for an overview. Those games are played on \textit{weighted} graphs, where edges are fitted with integer weights modeling rewards or costs. The performance of a play is evaluated via a \textit{payoff function} that maps it to the numerical domain. The objective of player~1 is then to ensure a sufficient payoff with regard to a given threshold value. Seminal classes of quantitative games include mean-payoff~($\MPG$), total-payoff~($\TPG$) and energy games~($\EG$). In~$\MPG$ games~\cite{EM79,ZP96,ipl68(3)-Jur}, player~1 has to optimize his long-run average gain per edge taken whereas, in $\TPG$ games~\cite{mfcs2004-GZ,GS09}, player~1 has to optimize his long-run sum of weights. Energy games~\cite{emsoft2003-CAHS,BFLMS08,JLR13} model safety-like properties: the goal is to ensure that the running sum of weights never drops below zero and/or that it never exceeds a given upper bound $U \in \mathbb{N}$. All three classes share common properties. First, $\MPG$~games, $\TPG$ games, and $\EG$ games with only a lower bound ($\EGL$) are memoryless determined (given an initial state, either player~1 has a strategy to win, or player~2 has one, and in both cases no memory is required to win). Second, deciding the winner for those games is in \NP $\cap$ \coNP and no polynomial algorithm is known despite many efforts (e.g.,~\cite{BCDGR11,Chatterjee201525}). Energy games with both lower and upper bounds~($\EGLU$) are more complex: they are \EXPTIME-complete and winning requires memory in general~\cite{BFLMS08}.

While those classes are well-known, it is sometimes necessary to go beyond them to accurately model practical applications. For example, multi-dimensional games and conjunctions with a parity objective model trade-offs between different quantitative aspects~\cite{CD10,CRR14,VCDHRR15}. Similarly, window objectives address the need for strategies ensuring good quantitative behaviors within reasonable time frames~\cite{Chatterjee201525}.

\paragraph{Average-energy games.} We study the \textit{average-energy} ($\AE$) payoff function: in $\AE$ games, the goal of player~1 is to optimize the \textit{long-run average accumulated energy} over a play. We introduce this objective to formalize the specification desired in a practical application~\cite{CJLRR09}, which we detail in the following as a motivating example. Interestingly, it turns out that this payoff first appeared long ago~\cite{TV87}, but it was not subject to a systematic study until very recently: see related work for more discussion.

In addition to being meaningful w.r.t.~practical applications, $\AE$ games also have theoretical interest. In~\cite{CP13}, Chatterjee and Prabhu define the \textit{average debit-sum level} objective, which can be seen as a variation of the \textit{average-energy} where the accumulated energy is taken to be zero in any point where it is actually positive (hence, it focuses on the average debt). They use the corresponding games to compute the values of quantitative timed simulation functions. In particular, they provide a pseudo-polynomial-time algorithm to solve those games, but the complexity of deciding the winner as well as the memory requirements are open. Here, we solve those questions for the very similar average-energy objective.

\paragraph{Motivating example.} Our example is a simplified version of the
industrial application studied by Cassez \textit{et~al.}~\cite{CJLRR09}. Consider a
machine that consumes oil, stored in a connected accumulator. We want to
synthesize an appropriate controller to operate the oil pump that fills the
accumulator, and by the effect of pressure, that releases oil from the accumulator into the machine with a (time-varying) rate according to desired production. In order to ensure safety, the oil level in the accumulator should be maintained at all times between a minimal and a maximal level. This part of the specification can be encoded as an energy objective with both lower and upper bounds~($\EGLU$). At the same time, the more oil (thus pressure) in the accumulator, the faster the whole apparatus wears out. Hence, an ideal controller should minimize the average level of oil in the long run. This desire can be formalized through the average-energy payoff~($\AE$). Overall, the specification is thus to minimize the average-energy under the strong energy constraints: we denote the corresponding objective by~$\AELU$.


\begin{table*}[thb]\centering
\small
\ra{1.1}
\scalebox{1}{\begin{tabular}{cccccc}\toprule
Game objective & \textbf{1-player} && \textbf{2-player}&& \textbf{memory}  \\
 \midrule

$\MPG$  & in \PTIME~\cite{Kar78} && in \NP $\cap$ \coNP~\cite{ZP96} && memoryless~\cite{EM79} \\
$\TPG$  & in \PTIME~\cite{FV97} && in \NP $\cap$ \coNP~\cite{GS09}  && memoryless~\cite{mfcs2004-GZ} \\
$\EGL$  & in \PTIME \cite{BFLMS08} && in \NP $\cap$ \coNP \cite{emsoft2003-CAHS,BFLMS08} && memoryless~\cite{emsoft2003-CAHS} \\
$\EGLU$ & \PSPACE-complete \cite{FJ13} && \EXPTIME-complete \cite{BFLMS08} && pseudo-polynomial \\
 \midrule

%

$\AEG$  & in \PTIME && in \NP $\cap$ \coNP && memoryless \\
$\AELU$, polynomial~$U$ & in \PTIME && in \NP $\cap$ \coNP && polynomial \\
$\AELU$, arbitrary~$U$ & in \EXPTIME~/~PSPACE-hard && \EXPTIME-complete && pseudo-polynomial \\
$\AEL$  & \EXPTIME-easy~/~\NP-hard && \textit{open}~/~\EXPTIME-hard && \textit{open} ($\geq$ pseudo-p.)\\
\bottomrule
\end{tabular}}
\vspace{1mm}

\caption{Complexity of deciding the winner and memory requirements for quantitative games: $\MP$~stands for mean-payoff, $\TP$ for total-payoff, $\EGL$ (resp.~$\EGLU$) for lower-bounded (resp.~lower- and upper-bounded) energy, $\AE$ for average-energy, and $\AEL$ (resp.~$\AELU$) for average-energy under a lower bound (resp.~and upper bound $U \in \mathbb{N}$) on the energy. Results without reference are proved in this paper.}
\label{tab:results}
\end{table*}

\paragraph{Contributions.} Our main results are summarized in Table~\ref{tab:results}.

A) We establish that the average-energy objective can be seen as a \textit{refinement} of total-payoff, in the same sense as total-payoff is seen as a refinement of mean-payoff~\cite{GS09}: it allows to distinguish strategies yielding identical mean-payoff and total-payoff. 

B) We show that deciding the winner in two-player $\AE$ games is in \NP $\cap$ \coNP whereas it is in \PTIME for one-player games. In both cases, memoryless strategies suffice (Thm.~\ref{thm:ae_two_memoryless}). Those complexities match the state-of-the-art for $\MP$ and $\TP$ games~\cite{ZP96,ipl68(3)-Jur,GS09,BCDGR11}. Furthermore we prove that $\AE$ games are at least as hard as mean-payoff games (Thm.~\ref{thm:mp_to_ae}). Therefore, the \NP $\cap$ \coNP-membership can be considered optimal w.r.t.~our knowledge of $\MP$ games.
Technically, the crux of our approach is as follows. First, we show that memoryless strategies suffice in one-player $\AE$ games (Thm.~\ref{thm:aeg_memoryless}): this requires to prove important properties of the $\AE$ payoff as classical sufficient criteria for memoryless determinacy present in the literature fail to apply directly. Second, we establish a polynomial-time algorithm for the one-player case: it exploits the structure of winning strategies and mixes graph techniques with local linear program solving (Thm.~\ref{thm:ae_onePlayer_PTIME}). Finally, we lift memoryless determinacy to the two-player case using results by Gimbert and Zielonka~\cite{GZ05} and obtain the \NP~$\cap$~\coNP-membership as a corollary (Thm.~\ref{thm:ae_npinter}). 
\enlargethispage{3mm}

C) We establish an \EXPTIME algorithm to solve two-player $\AE$ games with lower- and upper-bounded energy ($\AELU$) with an arbitrary upper bound $U \in \mathbb{N}$ (Thm.~\ref{thm:aelu_reduc}). It relies on a reduction of the $\AELU$ game to a pseudo-polynomially larger $\AE$ game where the energy constraints are encoded in the graph structure. Applying straightforwardly the $\AE$ algorithm on this game would only give us \NEXPTIME~$\cap$~\coNEXPTIME-membership, hence we avoid this blowup by further reducing the problem to a particular $\MP$ game and applying a pseudo-polynomial algorithm, with some care to ensure that overall the algorithm only requires pseudo-polynomial time in the original $\AELU$ game. Since the simpler $\EGLU$ games (i.e., $\AELU$ with a trivial $\AE$ constraint) are already \EXPTIME-hard~\cite{BFLMS08}, the $\EXPTIME$-membership result is optimal. We also prove that pseudo-polynomial memory is both sufficient and in general necessary to win in $\AELU$ games, for both players (Thm.~\ref{thm:aelu_memory}). Whether one-player $\AELU$ games belong to $\PSPACE$ is an open question. For polynomial (in the size of the game graph) values of the upper bound~$U$---or~if it is given in unary---the~complexity of the two-player $\AELU$ problem collapses to \NP $\cap$ \coNP with the same approach, and polynomial memory suffices for both players.

D) We provide partial answers for the $\AEL$ objective---$\AE$ under a lower
bound constraint on energy but no upper bound. We~provide an \EXPTIME
algorithm for the one-player case, by reducing the problem to an $\AELU$ game
with a sufficiently large upper bound. That is, we prove that if the player
can win for the $\AEL$ objective, then he can do so without ever increasing
its energy above a well-chosen bound. We also prove the $\AEL$ problem to be
at least \NP-hard in one-player games and \EXPTIME-hard in two-player games
(Lem.~\ref{lem:ael_exp_hard}) via reductions from the subset-sum problem and
countdown games respectively. Finally, we show that memory is required for
both players in two-player $\AEL$ games (Lem.~\ref{lem:ael_memory}), and that
pseudo-polynomial memory is both sufficient and necessary in the one-player
case (Thm.~\ref{thm:ael_one_memory}). The~decidability status of two-player $\AEL$ games remains open as we only provide a correct but incomplete incremental algorithm (Lem.~\ref{lem:ael_semi}). We conjecture that the two-player $\AEL$ problem is decidable and sketch a potential approach to solve it. We highlight the key remaining questions and discuss some connections with related models that are known to be difficult. 

Observe that in many applications, the energy must be stocked in a finite-capacity storage for which an upper bound is provided. Hence, the model of choice in this case is~$\AELU$.

\paragraph{Related work.} The \textit{average-energy} payoff---Eq.~\eqref{eq:ae}---appeared in a paper by
Thuijsman and Vrieze in the late eighties~\cite{TV87}, under the name
\textit{total-reward}. This definition is different from the classical
\textit{total-payoff}---see~Sect.~\ref{sec:prelim}---commonly studied in the formal
methods community (see for example~\cite{mfcs2004-GZ,GS09}), which, despite
that, has been referred in many papers as either total-payoff or total-reward
equivalently. We will see in this paper that both definitions are
\textit{indeed} different and exhibit different behaviors.

Maybe due to this confusion, the payoff of Eq.~\eqref{eq:ae}---which we call
\textit{average-energy} thus avoiding misunderstandings---was not studied
extensively until recently. Nothing was known about memoryless determinacy and
complexity of deciding the winner. Independently to our work, Boros et
al.~recently studied the same payoff (under the name \textit{total-payoff}).
In~\cite{BEGM15}, they study Markov decision processes and stochastic
games with the payoff of Eq.~\eqref{eq:ae} and solve both questions. Their
results overlap with ours for $\AE$ games (Table~\ref{tab:results}). Let us
first mention that our results were obtained independently. Second, and
\textit{most importantly}, our approach and \textit{techniques are different},
and we believe our take on the problem yields some interest for our community.
Indeed, the algorithm of Boros \textit{et~al.} entirely relies on linear programming in
the one-player case, and resorts to approximation by discounted games in the
two-player one. Our techniques are arguably more constructive and based on
inherent properties of the payoff. In that sense, it is closer to what is
usually deemed important in our field. For example, we provide an extensive
comparison with classical payoffs. We base our proof of memoryless determinacy
on \textit{operational understanding} of the $\AE$ which is crucial in order
to formalize proper specifications. Our technique then benefits from seminal
works~\cite{GZ05} to bypass the reduction to discounted games and obtain a
direct proof, thanks to our more constructive approach. Lastly,
while~\cite{BEGM15} considers the $\AEG$ problem in the stochastic
context, we focus on the deterministic one but consider multi-criteria
extensions by adding bounds on the energy ($\AELU$ and $\AEL$ games). Those
extensions are \textit{completely new}, exhibit theoretical interest and are
adequate for practical applications in constrained energy systems, as
witnessed by the case study of~\cite{CJLRR09}.

Recent work of Br\'azdil \textit{et~al.}~\cite{BKKN14} considers the optimization of a payoff under energy constraint. They study mean-payoff in consumption systems, i.e., simplified one-player energy games where all edges consume energy but some states can atomically produce a reload of the energy up to the allowed capacity.

Full details and proofs of the results presented here can be found in the extended paper~\cite{BMRLL15a}.

\section{Preliminaries}
\label{sec:prelim}

\paragraph{Graph games.} 
We consider turn-based games played on graphs between two players denoted by
$\pI$ and~$\pII$. A~\emph{game} is a tuple $\Game =
(S_1, S_2, \trans, \weg)$ where 
(i)~$S_1$~and $S_2$ are disjoint finite sets of
\textit{states} belonging to $\pI$ and~$\pII$, with $S = S_1
\uplus S_2$, 
(ii)~$\trans \subseteq S \times S$ is a finite set of \textit{edges}, and 
(iii)~$\weg\colon \trans \to \bbZ$ is an integer \textit{weight
  function}. 
Given edge $(s_{1}, s_{2}) \in \trans$, we write $\weg(s_{1},
s_{2})$ as a shortcut for $\weg((s_{1}, s_{2}))$. We denote by~$\largestW$ the
largest absolute weight assigned by function $\weg$. A~game is
called $1$-player if~$\states_{1} = \emptyset$ or $\states_{2} = \emptyset$.

A \emph{play} from an initial state $\initState \in \states$ is an infinite sequence $\play = s_0 s_1 \ldots s_n \ldots$ such
that $s_0 = \initState$ and for all $i \ge 0$ we have $(s_i,s_{i+1}) \in \trans$.
The (finite) \emph{prefix} of $\play$ up to position $n$ gives the sequence $\play(n) = s_0
s_1 \ldots s_n$, the last element $s_n$ is denoted $\last(\play(n))$. The set
of all plays in $\Game$ is denoted by $\plays(\Game)$ and the set of all
prefixes is denoted by $\prefs(\Game)$. We say that
a prefix $\prefix \in \prefs(\Game)$ belongs to $\player{i}$, $i \in \{1,2\}$, if $\last(\prefix) \in S_i$. The set of prefixes that belong to $\player{i}$ is denoted by $\prefs_i (\Game)$. The classical concatenation between prefixes (resp. prefix and play) is denoted by the $\cdot$ operator. The length of a non-empty prefix $\prefix = s_{0}\ldots{}s_{n}$ is defined as the number of edges and denoted by $\vert\prefix\vert = n$.

\paragraph{Payoffs of plays.} Given a play $\play = s_0 s_1 \ldots s_n \ldots$ we define
\begin{itemize}
\item its \textit{energy level} at position $n$ as
	$\EL(\play(n)) = \sum_{i = 0}^{n-1} w(s_i,s_{i+1})$;

\item its \textit{mean-payoff} as
	$\MPsup(\play) = 
	\limsup_{n \to \infty} \frac{1}{n}  \sum_{i = 0}^{n-1} w(s_i,s_{i+1}) 
	= \limsup_{n \to \infty} \frac{1}{n} \EL(\play(n))$;
	
\item its \textit{total-payoff} as
	$\TPsup(\play) = 
	\limsup_{n \to \infty} \sum_{i = 0}^{n-1} w(s_i,s_{i+1}) 
	= \limsup_{n \to \infty} \EL(\play(n))$;
\item and its \textit{average-energy} as
\vspace{-1mm}
\begin{equation}
\label{eq:ae}
  	\AEsup(\play) = 
	\limsup_{n \to \infty} \frac{1}{n} \sum_{i = 1}^{n} 
	\left ( \sum_{j = 0}^{i-1} w(s_j,s_{j+1}) \right )
	= \limsup_{n \to \infty} \frac{1}{n} \sum_{i = 1}^{n}  \EL(\play(i)).
\end{equation}
\end{itemize}
\vspace{-1mm}

We will sometimes consider those measures defined with $\liminf$ instead of $\limsup$, in which case we write $\MPinf$, $\TPinf$ and $\AEinf$ respectively. Finally, we also consider those measures over prefixes: we naturally define
them by dropping the $\limsup_{n \rightarrow \infty}$ operator and taking $n =
\vert\prefix\vert$ for a prefix $\prefix \in \prefs(\Game)$. In this case, we
simply write $\MP(\prefix)$, $\TP(\prefix)$ and $\AE(\prefix)$ to denote the
fact that we consider \textit{finite} sequences.

\paragraph{Strategies.} A~\emph{strategy} for $\player{i}$, $i \in \{1,2\}$, is a function $\St_i\colon \prefs_i(\Game)
\to S$ such that for all $\prefix \in \prefs_i(\Game)$ we have $(\last(\prefix),
\St_i(\prefix)) \in \trans$.
A~strategy~$\St_{i}$ for~$\player{i}$ is \textit{finite-memory} 
if it can be
encoded by a deterministic finite-state Moore machine. A strategy is \emph{memoryless} if it does not depend on the history but only on the current
state of the game. We denote by $\strats_{i}(\Game)$, the sets of strategies for player $\player{i}$. We drop $\Game$ when the context is clear.

A play $\play = s_0 s_1 \ldots$ is \emph{consistent} with a
strategy $\St_i$ of $\player{i}$ if, for all $n \ge 0$ where $\last(\play(n))
\in S_i$, we have $\St_i(\play(n)) = s_{n+1}$. Given an initial state
$\initState \in \states$ and strategies $\St_{1}$ and $\St_{2}$ for the two
players, we denote by $\out(\initState, \St_1,\St_2 )$ the unique play that
starts in $\initState$ and is consistent with both $\St_1$ and $\St_2$. When
fixing the strategy of only $\player{i}$, we denote the set of
consistent outcomes by $\outs(\initState, \St_i)$.

\paragraph{Objectives.} 
An~\emph{objective} in $\Game$ is a set $\mathcal{W} \subseteq \plays(\Game)$ that is declared
winning for $\pI$. Given a game $\Game$, an initial state~$\initState$, and an
objective~$\mathcal{W}$, a~strategy $\St_1 \in \strats_{1}$ is winning for $\playerOne$
if for all strategy $\St_2 \in \strats_{2}$, we have that $\out(\initState,
\St_1,\St_2) \in \mathcal{W}$. Symmetrically, a~strategy $\St_2 \in \strats_{2}$ is
winning for $\playerTwo$ if for all strategy $\St_1 \in \strats_{1}$, we~have
that $\out(\initState, \St_1,\St_2) \not\in \mathcal{W}$. That is, we consider \textit{zero-sum} games.

We consider the following objectives and combinations of those objectives.
\begin{itemize}
	\item Given an initial energy level $\initCredit \in \bbN$, the \textbf{lower-bounded energy} ($\EGL$) objective
		$\LBound(\initCredit) = \{ \play \in \plays(G)$ $\mid \forall\, n
		 \ge 0,\ \initCredit + \EL(\play(n)) \geq 0 \}$
	requires non-negative energy at all times. 
	
	\item Given an upper bound $U \in \bbN$ and an initial energy level
	$\initCredit \in \bbN$, the \textbf{lower- and upper-bounded energy} ($\EGLU$) objective
			$\LUBound(U, \initCredit) = \{ \play \in \plays(\Game) \mid 
			\forall\, n \ge 0,\ \initCredit + \EL(\play(n)) \in [0,U] \}$
		requires that the energy always remains non-negative and below the 
		upper bound $U$ along a play.

	\item Given a threshold $t \in \bbQ$, the \textbf{mean-payoff} ($\MPG$) objective 
		$\MeanPayOff(t) = \{ \play \in \plays(\Game) \mid \MPsup(\play) \le t \}$
	requires that the mean-payoff is at most~$t$. 
	
	\item Given a threshold $t \in \bbZ$, the \textbf{total-payoff} ($\TPG$) objective 
		$\TotalPayOff(t) = \{ \play \in \plays(\Game) \mid \TPsup(\play) \le t \}$
	requires that the total-payoff is at most~$t$.

	\item Given a threshold $t \in \bbQ$, the \textbf{average-energy} ($\AEG$) objective
		$\AvgEnergyLevel(t) = \{ \play \in \plays(\Game) \mid \AEsup(\play) \le t \}$
	requires that the average-energy is at most~$t$.
\end{itemize}

For the $\MPG$, $\TPG$ and $\AEG$ objectives, note that
$\playerOne$ aims to \textit{minimize} the payoff value while $\playerTwo$
tries to maximize~it. The reversed convention is also often used in the
literature but both are equivalent. For our motivating example, 
seeing $\playerOne$ as a minimizer is more natural. Note that we define the objectives using the $\limsup$ variants
of~$\MPG$, $\TPG$ and~$\AEG$, but similar results are obtained for the $\liminf$ variants.

\paragraph{Decision problem.} 
Given a game $\Game$, an initial state $\initState \in \states$, and an
objective $\mathcal{W} \subseteq \plays(\Game)$ as defined above, the associated
\textit{decision problem} is to decide if $\pI$ has a winning strategy for
this objective.

We recall classical results in Table~\ref{tab:results}. Memoryless strategies suffice for both players for $\EGL$~\cite{emsoft2003-CAHS,BFLMS08}, $\MPG$~\cite{EM79} and
$\TPG$~\cite{FV97,mfcs2004-GZ} objectives. Since all associated
problems can be solved in polynomial time for 1-player games, it follows that
the 2-player decision problem is in $\NP\cap\coNP$ for those three objectives~\cite{BFLMS08,ZP96,GS09}. For the
$\EGLU$ objective, memory is in general needed and the associated
decision problem is \EXPTIME-complete~\cite{BFLMS08} (\PSPACE-complete for one-player games~\cite{FJ13}).

\paragraph{Game values.} Given a game with an objective $\mathcal{W} \in \{\MeanPayOff, \TotalPayOff, \AvgEnergyLevel\}$ and an initial state $\initState$, we refer to the \textit{value} from $\initState$ as $v = \inf \{t \in \mathbb{Q} \mid \exists\, \sigma_{1} \in \Sigma_{1},\, \outs(\initState, \St_1) \subseteq \mathcal{W}(t)\}$. For~both $\MPG$ and~$\TPG$ objectives, it~is known that the value can be achieved by an optimal memoryless strategy; for the $\AEG$ objective it follows from our results (Thm.~\ref{thm:ae_two_memoryless}).

\section{Average-Energy}
\label{sec:average_games}

In this section, we consider the problem of ensuring a \textit{sufficiently low} average-energy.
\begin{bproblem}[$\AEG$] 
  Given a game~$\Game$, an initial state $\initState$, and a threshold~$t \in \mathbb{Q}$, decide if $\pI$ has a winning strategy $\St_1 \in \strats_{1}$ for the objective
  $\AvgEnergyLevel(t)$.
\end{bproblem}

\subsection{Relation with classical objectives}
\label{subsec:ae_relation}

Several links between $\EGL$, $\MPG$
and $\TPG$ objectives can be established. Intuitively, $\pI$ can only ensure a lower bound on energy if he can prevent $\pII$ from enforcing strictly-negative cycles (otherwise the initial energy is eventually exhausted). This is the case if and only if $\pI$ can ensure a non-negative mean-payoff in $\Game$ (here, he wants to maximize the $\MP$), and if this is the case, $\pI$ can prevent the running sum of weights from ever going too far beyond zero along a play, hence granting a lower bound on total-payoff.

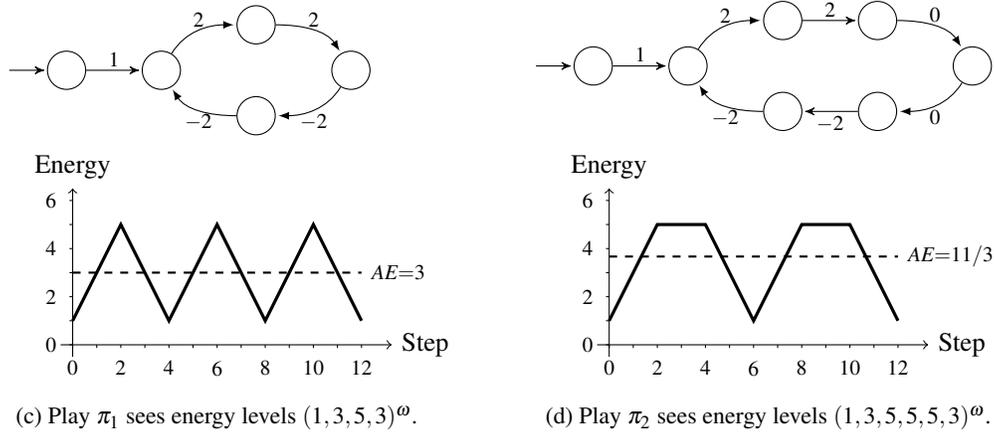
\begin{figure}[tb]
        \centering
\subfloat{\scalebox{1}{\begin{tikzpicture}[->,>=stealth',shorten >=1pt,auto,node
    distance=2.5cm,bend angle=45, scale=0.42, yscale=.9,font=\normalsize,inner sep=.5mm]
    \everymath{\scriptstyle}
    \tikzstyle{p1}=[draw,circle,text centered,minimum size=5mm,text width=4mm]
    \tikzstyle{p2}=[draw,rectangle,text centered,minimum size=5mm,text width=4mm]
    \node[p1]  (0)  at (0, 0) {};
    \node[p1]  (1) at (3, 0) {};
    \node[p1]  (2) at (6, 1.6) {};
    \node[p1]  (3) at (6, -1.6) {};
    \node[p1]  (4) at (9, 0) {};
    
    \coordinate[shift={(-5mm,0mm)}] (init) at (0.west);
    \path
    (0) edge node[above] {$1$} (1)
    (init) edge (0);
	\draw[->,>=latex] (1) to[out=60,in=180] node[above] {$2$} (2);
	\draw[->,>=latex] (2) to[out=0,in=120] node[above] {$2$} (4);
	\draw[->,>=latex] (4) to[out=240,in=0] node[below] {$-2$} (3);
	\draw[->,>=latex] (3) to[out=180,in=300] node[below] {$-2$} (1);
      \end{tikzpicture}} }
      \hspace{2cm}
      \subfloat{\scalebox{1}{\begin{tikzpicture}[->,>=stealth',shorten >=1pt,auto,node
    distance=2.5cm,bend angle=45, scale=0.42, yscale=.9, font=\normalsize,inner sep=.5mm]
    \everymath{\scriptstyle}
    \tikzstyle{p1}=[draw,circle,text centered,minimum size=5mm,text width=4mm]
    \tikzstyle{p2}=[draw,rectangle,text centered,minimum size=5mm,text width=4mm]
    \node[p1]  (0)  at (0, 0) {};
    \node[p1]  (1) at (3, 0) {};
    \node[p1]  (2) at (6, 1.6) {};
    \node[p1]  (3) at (6, -1.6) {};
    \node[p1]  (2b) at (9, 1.6) {};
    \node[p1]  (3b) at (9, -1.6) {};
    \node[p1]  (4) at (12, 0) {};
    
    \coordinate[shift={(-5mm,0mm)}] (init) at (0.west);
    \path
    (2) edge node[above] {$2$} (2b)
    (3b) edge node[below] {$-2$} (3)
    (0) edge node[above] {$1$} (1)
    (init) edge (0);
	\draw[->,>=latex] (1) to[out=60,in=180] node[above] {$2$} (2);
	\draw[->,>=latex] (2b) to[out=0,in=120] node[above] {$0$} (4);
	\draw[->,>=latex] (4) to[out=240,in=0] node[below] {$0$} (3b);
	\draw[->,>=latex] (3) to[out=180,in=300] node[below] {$-2$} (1);
      \end{tikzpicture}} }

\subfloat[Play $\play_{1}$ sees energy levels $(1, 3, 5, 3)^{\omega}$.]{\scalebox{1}{\begin{tikzpicture}[scale=.8]
      \everymath{\scriptstyle}
		\def \xscale {0.4}
		\def \yscale {0.4}
		\def \xmax {12}
		\def \ymax {6}
			
		\def\xy#1#2{({#1*\xscale},{#2*\yscale})}
                \path[use as bounding box] (-1,-.8) -- ({\xmax*\xscale+1},{\ymax*\yscale + 0.6});
	    \draw[->] (-0.2,0) -- ({\xmax*\xscale+0.5},0) node[right] {\small Step};
	    \draw[->] (0,-0.2) -- (0,{\ymax*\yscale + 0.2}) node[above] {\small Energy};
 		
        \foreach \y in {0,2,...,\ymax} {	
			\draw (-0.1,{\y*\yscale}) node[anchor=east] {$\y$} ;
		}
		
        \foreach \x in {0,2,...,\xmax} {
			\draw ({\x*\xscale},-0.1) node[anchor=north] {$\x$} ;
		}

        \foreach \x in {0,...,\xmax} {
            \draw ({\x*\xscale},0) -- ({\x*\xscale},-1.5pt);
        }

        \foreach \y in {0,...,\ymax} {
            \draw (0, {\y*\yscale}) -- (-1.5pt, {\y*\yscale});
        }
		
		\def\mean{3}
		\draw[thick,dashed,color=black]
			\xy{0}{\mean} -- \xy{\xmax}{\mean};
		
		\draw \xy{\xmax}{\mean} node[anchor=west] {$ \mathit{AE} = 3$ };

		\draw[very thick,-,color=black]
					\xy{0}{1} --
					\xy{2}{5} --
					\xy{4}{1} --
					\xy{6}{5} --
					\xy{8}{1} --
					\xy{10}{5} --
					\xy{12}{1}  ;							 
	\end{tikzpicture}}}
	\hspace{1.5cm}
\subfloat[Play $\play_{2}$ sees energy levels $(1, 3, 5, 5, 5, 3)^{\omega}$.]{\scalebox{1}{
\begin{tikzpicture}[scale=.8]
					
      \everymath{\scriptstyle}
		\def \xscale {0.4}
		\def \yscale {0.4}
		\def \xmax {12}
		\def \ymax {6}
			
		\def\xy#1#2{({#1*\xscale},{#2*\yscale})}
               \path[use as bounding box] (-1,-.8) -- ({\xmax*\xscale+1.5},{\ymax*\yscale + 0.6});
	    \draw[->] (-0.2,0) -- ({\xmax*\xscale+0.5},0) node[right] {\small Step};
	    \draw[->] (0,-0.2) -- (0,{\ymax*\yscale + 0.2}) node[above]
              {\small Energy};
 		
        \foreach \y in {0,2,...,\ymax} {	
			\draw (-0.1,{\y*\yscale}) node[anchor=east] {$\y$} ;
		}
		
        \foreach \x in {0,2,...,\xmax} {
			\draw ({\x*\xscale},-0.1) node[anchor=north] {$\x$} ;
		}

        \foreach \x in {0,...,\xmax} {
            \draw ({\x*\xscale},0) -- ({\x*\xscale},-1.5pt);
        }

        \foreach \y in {0,...,\ymax} {
            \draw (0, {\y*\yscale}) -- (-1.5pt, {\y*\yscale});
        }
 	
		\def\mean{3.67}
		\draw[thick,dashed,color=black]
			\xy{0}{\mean} -- \xy{\xmax}{\mean};
		
		\draw \xy{\xmax}{\mean} node[anchor=west] {$ \mathit{AE} = 11/3$ };
		
		\draw[very thick,-,color=black]
					\xy{0}{1} --
					\xy{2}{5} --
					\xy{4}{5} --
					\xy{6}{1} --
					\xy{8}{5} --
					\xy{10}{5} --
					\xy{12}{1} ;		
	\end{tikzpicture}
}}
	\caption{Both plays have identical mean-payoff and total-payoff: $\MPsup(\play_{1}) = \MPinf(\play_{1}) = \MPsup(\play_{2}) = \MPinf(\play_{2}) = 0$, $\TPsup(\play_{1}) = \TPsup(\play_{2}) = 5$, and $\TPinf(\play_{1}) = \TPinf(\play_{2}) = 1$. But play $\play_{1}$ has a lower average-energy: $\AEsup(\play_{1}) = \AEinf(\play_{1}) =  3 < \AEsup(\play_{2}) = \AEinf(\play_{2}) = 11/3$.}
	\label{fig:ae_refines}
\end{figure}

The $\TPG$ objective is sometimes seen as a \textit{refinement} of
$\MPG$ for the case where~$\pI$---as~a minimizer---can ensure~$\MP$ equal
to zero but not lower, i.e., the $\MPG$ game has value zero~\cite{GS09}. Indeed, one may use
the $\TP$ to further discriminate between strategies that guarantee
$\MP$~zero. In~the same philosophy, the~average-energy can
help in distinguishing strategies that yield identical total-payoffs. See Fig.~\ref{fig:ae_refines}. The~$\AE$ values in both examples can be computed easily using the upcoming technical lemmas (Sect.~\ref{subsec:ae_tech}).

In these examples, the average-energy is clearly comprised between the infimum and supremum total-payoffs. This remains true for any play. In particular, \textit{if the mean-payoff value from a state is not zero}, its total-payoff value is infinite and the following holds: either $\playerOne$ can force $\AE$ equal to $-\infty$ or $\playerTwo$ can force $\AE$ equal to~$+\infty$.

\subsection{Useful properties of the average-energy}
\label{subsec:ae_tech}

\paragraph{Classical sufficient criteria.}

Various sufficient criteria---or connected approaches---to deduce memoryless determinacy appear in the literature~\cite{EM79,BSV04,AR14,mfcs2004-GZ,Kop06}. Unfortunately, they cannot be applied straight out of the box to the $\AE$ payoff. Intuitively, a common requirement is for winning
objectives to be closed under \textit{cyclic permutation} and under
\textit{concatenation}. Without further assumptions, the $\AE$
objective satisfies neither. Indeed, consider cycles represented by sequences
of \textit{weights} $\mathcal{C}_{1} = \{-1\}$, $\mathcal{C}_{2} = \{1\}$ and
$\mathcal{C}_{3} = \{1, -2\}$. We see that
$\AE(\mathcal{C}_{1}\mathcal{C}_{2}) = (-1 + 0)/2 = -1/2 <
\AE(\mathcal{C}_{2}\mathcal{C}_{1}) = (1 -0)/2 = 1/2$, hence $AE$ is not
closed under cyclic permutations. Intuitively, the order in which the weights
are seen \textit{does} matter, in contrast to most classical payoffs. For
concatenation, see that $\AE(\mathcal{C}_{3}) = 0$ while
$\AE(\mathcal{C}_{3}\mathcal{C}_{3}) = -1/2 < 0$. Here the intuition is that
the overall $\AE$ is impacted by the energy of the first cycle
which is strictly negative ($-1$). In a sense, the $\AE$ of a cycle
can only be maintained through repetition if this cycle is neutral with regard
to the total energy level, i.e., if it has energy level zero: we will formalize this
intuition in Lem.~\ref{lem:AE_repeat}.

\paragraph{Extraction of prefixes.} We establish two useful properties of the average-energy that help us to prove memoryless determinacy. The following lemma describes the impact of adding a finite prefix to an infinite play: it will help us in decomposing plays when needed.

\begin{restatable}{lemma}{lemAEprefix}[\textbf{Average-energy prefix}]
\label{lem:AE_prefix}
Let $\prefix \in \prefs(\Game)$, $\play \in \plays(\Game)$. Then, $\AEsup(\prefix \cdot \play) = \EL(\prefix) + \AEsup(\play)$. The same equality holds for $\AEinf$.
\end{restatable}

\paragraph{Extraction of a best cycle.} The next lemma is crucial to prove that memoryless strategies suffice: under well-chosen conditions, one can always select a best cycle in a play---hence, there is no interest in mixing different cycles and no use for memory. It holds only for sequences of cycles \textit{that have energy level zero}: since they do not change the energy, they do not modify the $\AE$ of the following suffix of play, and one can decompose the $\AE$ as a weighted average over zero cycles. The concatenation of cycles $\cycle_{a} = s\,s'\ldots{} s$ and $\cycle_{b} = s\,s'' \ldots{} s$ is to be understood as $\cycle_{a} \cdot \cycle_{b} = s\,s'\ldots{} s\,s'' \ldots{} s$.

\begin{restatable}{lemma}{lemAErepeat}[\textbf{Repeated zero cycles of bounded length}]
  \label{lem:AE_repeat}
  Let $\cycle_{1}, \cycle_{2}, \cycle_{3}, \ldots{}$ be an infinite sequence
  of cycles $\cycle_{i} \in \prefs(\Game)$ such that (i) $\play = \cycle_{1}
  \cdot \cycle_{2} \cdot \cycle_{3} \cdots{} \in \plays(\Game)$,
  (ii)
  $\forall\,i \geq 1$, $\EL(\cycle_{i}) = 0$ and (iii) $\exists\, \ell \in
  \bbN*$ such that $\forall\,i \geq 1$, $\vert\cycle_{i}\vert \leq \ell$. Then
  the following properties hold.
\begin{enumerate}
\item\label{prop:weighted_average} The average-energy of $\play$ is the
  \textit{weighted average} of the average-energies of the cycles:
\begin{equation}
  \AEsup(\play) = \limsup_{k \rightarrow \infty} \left[\dfrac{\sum_{i=1}^{k} 
      \vert\cycle_{i}\vert \cdot \AE(\cycle_{i})}{\sum_{i=1}^{k}
      \vert\cycle_{i}\vert}\right].
\end{equation}
\item\label{prop:repeated_cycle} For any cycle $\cycle \in \prefs(\Game)$ such
  that $\EL(\cycle) = 0$, we have that $\AEsup(\cycle^{\omega}) =
  \AE(\cycle)$.
\item\label{prop:repeat_best_cycle} Repeating the best cycle gives the lowest
  $\AE$:
  $\inf_{i \in \bbN*} \AE(\cycle_{i}) = \inf_{i \in \bbN*}
  \AEsup((\cycle_{i})^{\omega}) \leq \AEsup(\play)$.
\end{enumerate}
Similar properties hold for $\AEinf$.
\end{restatable}

\subsection{One-player games}

\textit{We assume that the unique player is $\playerOne$}, hence that $S_{2} = \emptyset$. The proofs are similar for the case where all states belong to $\playerTwo$ (i.e., $S_{1} = \emptyset$). Similarly, we present our results for the $\AEsup$ variant, but they carry over to the $\AEinf$ one. Actually, since we show that we can restrict ourselves to \textit{memoryless} strategies, all consistent outcomes will be periodic and thus both variants will be equal over those outcomes. 

\paragraph{Memoryless determinacy.} Intuitively, we use Lem.~\ref{lem:AE_prefix} and Lem.~\ref{lem:AE_repeat} to transform any arbitrary path in a simple \textit{lasso path}, repeating a unique simple cycle, and yielding an at least as good $\AE$, thus proving that any threshold achievable with memory can also be achieved without it.

\label{subsec:ae_one}
\begin{restatable}{theorem}{thmAEoneMemory}
\label{thm:aeg_memoryless}
Memoryless strategies are sufficient to win one-player $\AEG$ games.
\end{restatable}

\paragraph{Polynomial-time algorithm.} We know the form of optimal memoryless strategies: an optimal lasso path $\play = \prefix \cdot \calC^\omega$ w.r.t.~the $\AE$. We establish a polynomial-time algorithm to solve one-player $\AEG$ games. 

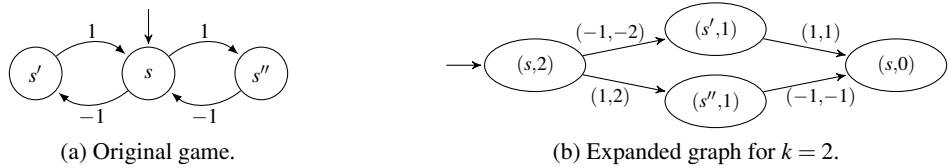
\begin{figure}[tb]
        \centering
\subfloat[Original game.]{\scalebox{1}{\begin{tikzpicture}[->,>=stealth',shorten >=1pt,auto,node
    distance=2.5cm,bend angle=45, scale=0.5, font=\normalsize,inner sep=.5mm]
    \everymath{\scriptstyle}
    \tikzstyle{p1}=[draw,circle,text centered,minimum size=7mm,text width=5mm]
    \tikzstyle{p2}=[draw,rectangle,text centered,minimum size=7mm,text width=4mm]
    \node[p1]  (0)  at (0, 0) {$s'$};
    \node[p1]  (1) at (3, 0) {$s$};
    \node[p1]  (2) at (6, 0) {$s''$};
    
    \coordinate[shift={(0mm,5mm)}] (init) at (1.north);
    \path
    (init) edge (1);
	\draw[->,>=latex] (0) to[out=40,in=140] node[above] {$1$} (1);
	\draw[->,>=latex] (1) to[out=40,in=140] node[above] {$1$} (2);
	\draw[->,>=latex] (2) to[out=220,in=-40] node[below] {$-1$} (1);
	\draw[->,>=latex] (1) to[out=220,in=-40] node[below] {$-1$} (0);
      \end{tikzpicture}}}
      \hspace{2cm}
      \subfloat[Expanded graph for $k = 2$.]{\scalebox{1}{\begin{tikzpicture}[->,>=stealth',shorten >=1pt,auto,node
    distance=2.5cm,bend angle=45, scale=.8, font=\small,inner sep=.5mm]
    \everymath{\scriptstyle}
    \tikzstyle{p1}=[draw,ellipse,text centered,minimum size=7mm,text width=8.5mm]
    \tikzstyle{p2}=[draw,rectangle,text centered,minimum size=7mm,text width=4mm]
    \node[p1]  (0)  at (0, 0) {$(s, 2)$};
    \node[p1]  (1) at (3, 0.6) {$(s', 1)$};
    \node[p1]  (2) at (3, -0.6) {$(s'', 1)$};
    \node[p1]  (4) at (6, 0) {$(s, 0)$};
    \coordinate[shift={(-5mm,0mm)}] (init) at (0.west);
    \path
    (0) edge node[above,xshift=-2mm] {$(-1, -2)$} (1)
    (0) edge node[below,xshift=-2mm] {$(1, 2)$} (2)
    (1) edge node[above,xshift=2mm] {$(1, 1)$} (4)
    (2) edge node[below,xshift=2mm] {$(-1, -1)$} (4)
    (init) edge (0);
      \end{tikzpicture}}}
	\caption{The best cycle $\calC_{s, 2}$ is computed by looking for a
          path from $(s,2)$ to $(s,0)$ with sum zero in the first dimension
          (zero cycle) and minimal sum in the second dimension (minimal
          $\AEG$). Here, the cycle via $s'$ is clearly better, with $\AEG$
          equal to $-1/2$ in contrast to $1/2$ via $s''$.}
	\label{fig:LPGraphForAE}
\end{figure}

\looseness=-1
The crux is
  computing, for each state~$s$, the best---w.r.t.~the~$\AE$---\textit{zero} cycle~$\calC_s$ starting and ending in~$s$ (if~any). This is achieved through linear programming~(LP) over expanded graphs. For~each state~$s$ and length $k \in \{1, \ldots{}, \vert \states \vert\}$, we compute the best cycle $\calC_{s, k}$ by considering a graph (Fig.~\ref{fig:LPGraphForAE}) that models all cycles of length~$k$ from~$s$ and that uses $k+1$ levels and two-dimensional weights on edges of the form $(c, l\cdot c)$ where $c$ is the weight in the original game and $l \in \{k, k-1, \ldots{}, 1\}$ is the level of the edge. In~the~LP, we~look for cycles $\calC_{s,k}$ of length $k$ on $s$ such that (a)~the~sum of weights in the first dimension is zero (thus $\calC_{s,k}$ is a \textit{zero} cycle), and (b)~the~sum in the second one is minimal. Fortunately, this sum is exactly equal to $\AE(\calC) \cdot k$ thanks to the $l$ factors used in the weights of the expanded graph. Hence, we obtain the optimal cycle $\calC_{s, k}$ (in~polynomial time).
Doing this $\vert \states \vert$ times for each state~$s$, we~obtain for each of them the optimal cycle $\calC_{s}$ (if one zero cycle exists). Then, by
  Lem.~\ref{lem:AE_prefix}, it~remains to compute the least~$\EL$
  with which each state~$s$ can be reached using classical graph techniques (e.g., Bellman-Ford), and to pick the optimal combination to obtain an optimal memoryless strategy, in polynomial time.

\begin{restatable}{theorem}{thmAEonePTIME}
\label{thm:ae_onePlayer_PTIME}
  The $\AEG$ problem for one-player games is in \PTIME.	
\end{restatable}

\subsection{Two-player games}
\label{subsec:ae_two}
\paragraph{Memoryless determinacy.} We now prove that memoryless strategies still suffice in two-player games. As discussed in Sect.~\ref{subsec:ae_tech}, classical criteria do not apply. There is, however, one result that proves particularly useful. Consider any payoff function such that memoryless strategies suffice for \textit{both} \textit{one-player} versions ($S_{1} = \emptyset$, resp. $S_{2} = \emptyset$). In~\cite[Cor.~7]{GZ05}, Gimbert and Zielonka establish that memoryless strategies also suffice in \textit{two-player} games with the same payoff.
\looseness=-1
Thanks to Thm.~\ref{thm:aeg_memoryless}, this entails the next theorem.

\begin{theorem}
\label{thm:ae_two_memoryless}
Average-energy games are determined and both players have memoryless optimal strategies.
\end{theorem}

\paragraph{Solving average-energy games.} By Thm.~\ref{thm:ae_two_memoryless}, one can guess an optimal memoryless strategy for $\playerTwo$ and solve the remaining one-player game for $\playerOne$, in polynomial time (by Thm.~\ref{thm:ae_onePlayer_PTIME}). The converse is also true: one can guess the strategy of $\playerOne$ and solve the remaining game where $S_{1} = \emptyset$ in polynomial time.

\begin{theorem}
\label{thm:ae_npinter}
The $\AEG$ problem for two-player games is in \NP $\cap$ \coNP.
\end{theorem}

We prove that $\MPG$ games can be encoded into $\AEG$ ones in polynomial time. The former are known to be in \NP $\cap$ \coNP but whether they belong to $\PTIME$ is a long-standing open question (e.g.,~\cite{ZP96,ipl68(3)-Jur,BCDGR11,Chatterjee201525}). Hence, w.r.t.~current knowledge, the \NP $\cap$ \coNP-membership of the $\AEG$ problem can be considered optimal. The~key of the construction is to double each edge of the original game and modify the weight function such that each pair of successive edges corresponding to such a doubled edge now has a total energy level of zero, and an average-energy that is exactly equal to the weight of the original edge. Then we apply decomposition techniques as in Lem.~\ref{lem:AE_repeat} to establish the equivalence.

\begin{restatable}{theorem}{thmMPtoAE}
\label{thm:mp_to_ae}
Mean-payoff games can be reduced to average-energy games in polynomial time.
\end{restatable}

\section{Average-Energy with Lower- and Upper-Bounded Energy}
\label{sec:average_lu}
We extend the $\AEG$ framework with constraints on the running energy level of the system. Such constraints are natural in many applications where the energy capacity is bounded (e.g., fuel tank, battery charge). We first study the case where the energy is subject to \textit{both} a lower bound (here, zero) \textit{and} an upper bound ($U \in \mathbb{N}$). 
We study the problem for the \textit{fixed initial energy level} $\initCredit \coloneqq 0$.
\begin{bproblem}[$\AELU$] 
Given a game~$\Game$, an initial state $\initState$, an upper bound $U \in \mathbb{N}$, and a threshold~$t \in \mathbb{Q}$, decide if $\pI$ has a winning strategy $\St_1 \in \strats_{1}$ for the objective $\LUBound(U, \initCredit \coloneqq 0)\,\cap\, \AvgEnergyLevel(t)$.
\end{bproblem}

\colorlet{newgreen}{green!60!black}

\begin{figure}[tb]
	\centering
	\subfloat[One-player $\AELU$ game.]{\label{fig:aelu_game}\scalebox{1}{\begin{tikzpicture}[->,>=stealth',shorten >=1pt,auto,node
    distance=2.5cm,bend angle=45, scale=0.6, font=\normalsize,scale=.75,inner sep=0.5mm]
    \everymath{\scriptstyle}
    \tikzstyle{p1}=[draw,circle,text centered,minimum size=7mm,text width=4mm]
    \tikzstyle{p2}=[draw,rectangle,text centered,minimum size=7mm,text width=4mm]
    \node[p1]  (0)  at (0, 0) {$b$};
    \node[p1]  (1) at (3, 0) {$a$};
    \node[p1]  (2) at (6, 0) {$c$};
    
    \coordinate[shift={(0mm,5mm)}] (init) at (1.north);
    \path
    (init) edge (1)
    (1) edge [loop below, out=240, in=300,looseness=2, distance=2cm] node [below] {$2$} (1);
	\draw[->,>=latex] (0) to[out=40,in=140] node[above] {$0$} (1);
	\draw[->,>=latex] (1) to[out=40,in=140] node[above] {$1$} (2);
	\draw[->,>=latex] (2) to[out=220,in=-40] node[below] {$0$} (1);
	\draw[->,>=latex] (1) to[out=220,in=-40] node[below] {$-3$} (0);
      \end{tikzpicture}}}	
	\subfloat[Play $\play_{1} = (acacacab)^{\omega}$.]{\scalebox{0.68}{
	\begin{tikzpicture}

		\def \xscale {0.6}
		\def \yscale {0.6}
		\def \xmax {8}
		\def \ymax {3}
			
		\def\xy#1#2{({#1*\xscale},{#2*\yscale})}
			   
	    \draw[->] (-0.2,0) -- ({\xmax*\xscale+0.5},0) node[right] {Step};
	    \draw[->] (0,-0.2) -- (0,{\ymax*\yscale + 0.2}) node[above] {Energy};
 		
        \foreach \y in {0,1,...,\ymax} {	
			\draw (-0.1,{\y*\yscale}) node[anchor=east] {\y} ;
		}
		
        \foreach \x in {1,2,...,\xmax} {
			\draw ({\x*\xscale},-0.1) node[anchor=north] {\x} ;
		}

        \foreach \x in {0,...,\xmax} {
            \draw ({\x*\xscale},0) -- ({\x*\xscale},-1.5pt);
        }

        \foreach \y in {0,...,\ymax} {
            \draw (0, {\y*\yscale}) -- (-1.5pt, {\y*\yscale});
        }
		
		\def\mean{3/2}
		\draw[thick,dashed,color=black]
			\xy{0}{\mean} -- \xy{\xmax}{\mean};
		
		\draw \xy{\xmax}{\mean} node[anchor=west] {$ \mathit{AE} = 3/2$ };

		\draw[very thick,-,color=black]
					\xy{0}{0} --
					\xy{1}{1} --
					\xy{2}{1} --
					\xy{3}{2} --
					\xy{4}{2} --
					\xy{5}{3} --
					\xy{6}{3} --
					\xy{7}{0} --
					\xy{8}{0} ;
					 
	\end{tikzpicture}}}
	\subfloat[Play $\play_{2} = (aacab)^{\omega}$.]{\scalebox{0.68}{
	\begin{tikzpicture}

		\def \xscale {0.6}
		\def \yscale {0.6}
		\def \xmax {5}
		\def \ymax {3}
			
		\def\xy#1#2{({#1*\xscale},{#2*\yscale})}
			   
	    \draw[->] (-0.2,0) -- ({\xmax*\xscale+0.5},0) node[right] {Step};
	    \draw[->] (0,-0.2) -- (0,{\ymax*\yscale + 0.2}) node[above] {Energy};
 		
        \foreach \y in {0,1,...,\ymax} {	
			\draw (-0.1,{\y*\yscale}) node[anchor=east] {\y} ;
		}
		
        \foreach \x in {1,2,...,\xmax} {
			\draw ({\x*\xscale},-0.1) node[anchor=north] {\x} ;
		}

        \foreach \x in {0,...,\xmax} {
            \draw ({\x*\xscale},0) -- ({\x*\xscale},-1.5pt);
        }

        \foreach \y in {0,...,\ymax} {
            \draw (0, {\y*\yscale}) -- (-1.5pt, {\y*\yscale});
        }
		
		\def\mean{8/5}
		\draw[thick,dashed,color=black]
			\xy{0}{\mean} -- \xy{\xmax}{\mean};
		
		\draw \xy{\xmax}{\mean} node[anchor=west] {$ \mathit{AE} = 8/5$ };

		\draw[very thick,-,color=black]
					\xy{0}{0} --
					\xy{1}{2} --
					\xy{2}{3} --
					\xy{3}{3} --
					\xy{4}{0} --
					\xy{5}{0} ;
					 
	\end{tikzpicture}}}
	\subfloat[Play {\color{newgreen}$\play_{3} = (acaab)^{\omega}$}.]{\scalebox{0.68}{
	\begin{tikzpicture}

		\def \xscale {0.6}
		\def \yscale {0.6}
		\def \xmax {5}
		\def \ymax {3}
			
		\def\xy#1#2{({#1*\xscale},{#2*\yscale})}
			   
	    \draw[->] (-0.2,0) -- ({\xmax*\xscale+0.5},0) node[right] {Step};
	    \draw[->] (0,-0.2) -- (0,{\ymax*\yscale + 0.2}) node[above] {Energy};
 		
        \foreach \y in {0,1,...,\ymax} {	
			\draw (-0.1,{\y*\yscale}) node[anchor=east] {\y} ;
		}
		
        \foreach \x in {1,2,...,\xmax} {
			\draw ({\x*\xscale},-0.1) node[anchor=north] {\x} ;
		}

        \foreach \x in {0,...,\xmax} {
            \draw ({\x*\xscale},0) -- ({\x*\xscale},-1.5pt);
        }

        \foreach \y in {0,...,\ymax} {
            \draw (0, {\y*\yscale}) -- (-1.5pt, {\y*\yscale});
        }
		
		\def\mean{1}
		\draw[thick,dashed,color=black]
			\xy{0}{\mean} -- \xy{\xmax}{\mean};
		
		\draw \xy{\xmax}{\mean} node[anchor=west] {$ \mathit{AE} = 1$ };

		\draw[very thick,-,color=newgreen]
					\xy{0}{0} --
					\xy{1}{1} --
					\xy{2}{1} --
					\xy{3}{3} --
					\xy{4}{0} --
					\xy{5}{0} ;
					 
	\end{tikzpicture}}}
	\caption{Example of a one-player $\AELU$ game ($U = 3$) and the evolution of energy under different strategies that maintain it within $\left[0,\, 3\right]$ at all times. The minimal average-energy is obtained with {\color{newgreen}play $\play_{3}$}: alternating in order between the $+1$, $+2$ and $-3$ cycles.}\label{fig:aelu_example}	
\end{figure}
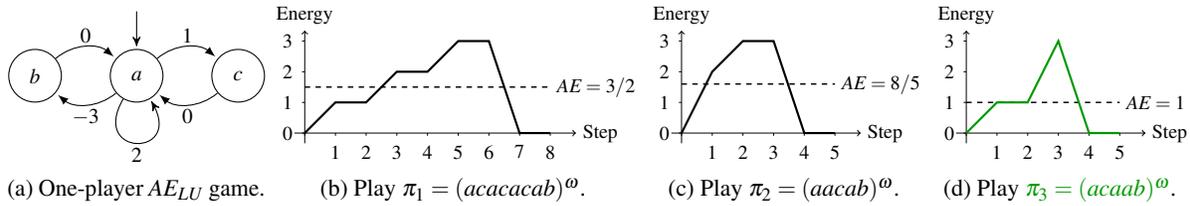

\paragraph{Illustration.} Consider the one-player game in Fig.~\ref{fig:aelu_example}. The energy constraints force $\playerOne$ to keep the energy in $[0,\,3]$ at all times. Hence, only three strategies can be followed safely, respectively inducing plays $\play_{1}$, $\play_{2}$ and $\play_{3}$. Due to the bounds on energy, it is natural that strategies need to alternate between both a positive and a negative cycle to satisfy objective $\LUBound(U, \initCredit \coloneqq 0)$ (since no simple zero cycle exists). It is yet interesting that to play optimally (play $\play_{3}$), $\playerOne$ actually has to use \textit{both} positive cycles, and in the \textit{appropriate order} (compare plays $\play_{2}$ and $\play_{3}$). 
\looseness=-1
This type of alternation is more intricate than for other classical objectives~\cite{CD10,CRR14,VCDHRR15}. This gives a hint of the complexity of $\AELU$ games.

\subsection{Pseudo-polynomial algorithm and complexity bounds}
\label{subsec:AELU_algo}

We first reduce the $\AELU$ problem to the $\AE$ problem over a \textit{pseudo-polynomial expanded game}, i.e., polynomial in the size of the original $\AELU$ game and in $U \in \mathbb{N}$. By Thm.~\ref{thm:ae_npinter} and Thm.~\ref{thm:ae_onePlayer_PTIME}, this reduction induces $\NEXPTIME \cap \coNEXPTIME$-membership of the two-player $\AELU$ problem, and $\EXPTIME$-membership of the one-player one. We improve the complexity for two-player games by further reducing the $\AE$ game to an $\MP$ game. This yields $\EXPTIME$-membership, which is optimal (Thm.~\ref{thm:aelu_reduc}).

Observe that if $U$ is encoded in unary or if~$U$ is polynomial in the size of the original game, the complexity of the $\AELU$ problem collapses to $\NP \cap \coNP$ for two-player games and to $\PTIME$ for one-player games thanks to our reduction to an $\AE$ problem and the results of Thm.~\ref{thm:ae_npinter} and Thm.~\ref{thm:ae_onePlayer_PTIME}.

\begin{figure}[tb]
\hspace{2.6cm}\scalebox{0.7}{\begin{tikzpicture}[->,>=stealth',shorten >=1pt,auto,node
    distance=2.5cm,bend angle=45, scale=1.2, font=\small]
    \tikzstyle{p1}=[draw,ellipse,text centered,minimum size=9mm,text width=8mm]
    \tikzstyle{p2}=[draw,rectangle,text centered,minimum size=7mm,text width=4mm]
    \node[p1]  (a0)  at (0, 0) {$(a, 0)$};
    \node[p1]  (a1) at (3, 0) {$(a, 1)$};
    \node[p1]  (a2) at (6, 0) {$(a, 2)$};
    \node[p1]  (a3) at (9, 0) {$(a, 3)$};
    \node[p1]  (b0)  at (-0.6, -1.2) {$(b, 0)$};
    \node[p1]  (b1) at (2.4, -1.2) {$(b, 1)$};
    \node[p1]  (b2) at (5.4, -1.2) {$(b, 2)$};
    \node[p1]  (b3) at (8.4, -1.2) {$(b, 3)$};
    \node[p1]  (c0)  at (-2, -2) {$(c, 0)$};
    \node[p1]  (c1) at (1, -2) {$(c, 1)$};
    \node[p1]  (c2) at (4, -2) {$(c, 2)$};
    \node[p1]  (c3) at (7, -2) {$(c, 3)$};
    \node[p1]  (sink) at (4.5, 1.9) {$\textsf{sink}$};
    \coordinate[shift={(-5mm,0mm)}] (init) at (a0.west);
    \path
    (a0) edge[thick,color=newgreen] node[right] {$1 \mid 0$} (c1)
    (a1) edge node[right] {$1 \mid 1$} (c2)
    (a2) edge node[right] {$1 \mid 2$} (c3)
    (b0) edge[thick,color=newgreen] node[right,yshift=-1mm,xshift=-1mm] {$0 \mid 0$} (a0)
    (b1) edge node[right,yshift=-1mm,xshift=-1mm] {$0 \mid 1$} (a1)
    (b2) edge node[right,yshift=-1mm,xshift=-1mm] {$0 \mid 2$} (a2)
    (b3) edge node[right,yshift=-1mm,xshift=-1mm] {$0 \mid 3$} (a3)
    (a0) edge node[left,xshift=-4mm] {$1 \mid 0$} (sink)
    (a1) edge node[left,xshift=-1mm] {$1 \mid 1$} (sink)
    (a2) edge node[right,xshift=1mm] {$1 \mid 2$} (sink)
    (a3) edge node[right,xshift=4mm] {$1 \mid 3$} (sink)
    (sink) edge [loop below, out=245, in=295,looseness=1, distance=0.8cm] node [below] {$1 \mid 2$} (sink)
    (init) edge (a0);
	\draw[->,>=latex] (c0) to[out=60,in=210] node[left,xshift=3mm,yshift=3mm] {$0 \mid 0$} (a0);
	\draw[->,>=latex,thick,color=newgreen] (c1) to[out=60,in=210] node[left,xshift=3mm,yshift=3mm] {$0 \mid 1$} (a1);
	\draw[->,>=latex] (c2) to[out=60,in=210] node[left,xshift=3mm,yshift=3mm] {$0 \mid 2$} (a2);
	\draw[->,>=latex] (c3) to[out=60,in=210] node[left,xshift=3mm,yshift=3mm] {$0 \mid 3$} (a3);
	\draw[->,>=latex,thick,color=newgreen] (a3) -- (10,0) -- (10,-2.5) to[out=180,in=0] node[above,xshift=8mm] {$-3 \mid 3$} (-0.6,-2.5) to[out=90, in=270] (b0);
	\draw[->,>=latex] (a0) to[out=15,in=165] node[above,xshift=-7mm,yshift=-1mm] {$2 \mid 0$} (a2);
	\draw[->,>=latex,thick,color=newgreen] (a1) to[out=15,in=165] node[above,xshift=7mm,yshift=-1mm] {$2 \mid 1$} (a3);
      \end{tikzpicture}}
	\caption{Reduction from the $\AELU$ game in Fig.~\ref{fig:aelu_game} to an $\AE$ game and further reduction to an $\MP$ game over the same expanded graph. For the sake of succinctness, the weights are written as $c \mid c'$ with $c$ the weight used in the $\AE$ game and $c'$ the one used in the $\MP$ game. We use the upper bound $U = 3$ and the average-energy threshold $t = 1$ (the optimal value in this case). The optimal play {\color{newgreen}$\play_{3} = (acaab)^{\omega}$} of the original game corresponds to an optimal memoryless play in the expanded graph.}\label{fig:aelu_to_ae}	
\vspace{-4mm}
\end{figure}
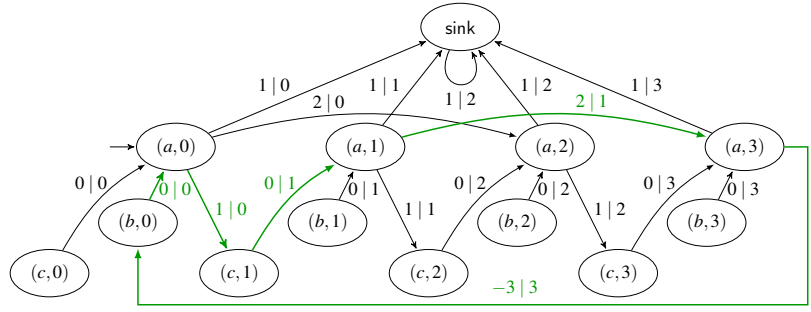

\paragraph{The reductions.} Given a game $G = (S_{1}, S_{2}, E, w)$, an initial state $\initState$, an upper bound $U \in \mathbb{N}$, and a threshold $t \in \mathbb{Q}$, we reduce the $\AELU$ problem to an $\AEG$ problem as follows. If at any point along a play, the energy drops below zero or exceeds $U$, the play will be losing for the $\LUBound(U, \initCredit \coloneqq 0)$ objective, hence also for its conjunction with the $\AEG$ one. So we build a new game $G'$ over the state space $(S \times \{0, 1, \ldots{}, U\}) \cup \{\textsf{sink}\}$. The idea is to include the energy level within the state labels, with $\textsf{sink}$ as an absorbing state reached only when the energy constraint is breached. We now consider the $\AEG$ problem for threshold $t$ on $G'$. By putting a self-loop of weight $1$ on $\textsf{sink}$, we ensure that if the energy constraint is not guaranteed in $G$, the answer to the $\AEG$ problem in $G'$ will be \textsf{No} as the average-energy will be infinite due to reaching this positive loop and repeating it forever.
Hence, we show that the $\AELU$ objective can be won in $G$ if and only if the $\AE$ one can be won in $G'$ (thus avoiding the $\textsf{sink}$ state). The result of the reduction for the game in Fig.~\ref{fig:aelu_game} is presented in Fig.~\ref{fig:aelu_to_ae}.

We then show that the $\AE$ game $G'$ can be further reduced to an $\MP$ game $G''$ by modifying the weight structure of the graph. Essentially, all edges leaving a state $(s, c)$ of $G'$ are given weight $c$ in $G''$, i.e., the current energy level, and the self-loop on \textsf{sink} is given weight $(\lceil t\rceil + 1)$. This modification is depicted in Fig.~\ref{fig:aelu_to_ae}. We claim that the $\AEG$ problem for threshold $t \in \mathbb{Q}$ in $G'$ is equivalent to the $\MPG$ problem for the same threshold in $G''$. Indeed, we show that with our change of weight function, reaching \textsf{sink} implies losing, both in $G'$ for $\AEG$ and in $G''$ for $\MPG$, and all plays that \textit{do not} reach \textsf{sink} have the same value for their average-energy in $G'$ as for their mean-payoff in $G''$.

\paragraph{Illustration.} Consider the $\AELU$ game $G$ in Fig.~\ref{fig:aelu_game}. The optimal strategy is $\play_{3} = (acaab)^{\omega}$. Now consider the reduction to the $\AE$ game, and further to the $\MP$ game, depicted in Fig.~\ref{fig:aelu_to_ae}. The optimal (memoryless) strategy in both the $\AE$ game $G'$ and the $\MPG$ game $G''$ is to create the play $\play' = ((a,0) (c,1) (a,1) (a,3) (b,0))^{\omega}$, which corresponds to the optimal play $\play_{3}$ in the original game. It can be checked that $\AEsup_{G}(\play_{3}) = \AEsup_{G'}(\play') = \MPsup_{G''}(\play')$. 

\paragraph{Complexity.} The reduction from the $\AELU$ game to the $\AE$ one induces a pseudo-polynomial blow-up in the number of states. Thanks to the second reduction and the use of a pseudo-polynomial algorithm for the $\MP$ game~\cite{ZP96,BCDGR11}, we get \EXPTIME-membership, which is optimal for two-player games thanks to the lower bound proved for $\EGLU$~\cite{BFLMS08}.

\begin{restatable}{theorem}{thmAELUcomplexity}
\label{thm:aelu_reduc}
The $\AELU$ problem is $\EXPTIME$-complete for two-player games and at least $\PSPACE$-hard for one-player games. If the upper bound $U \in \mathbb{N}$ is polynomial in the size of the game or encoded in unary, the $\AELU$ problem collapses to $\NP \cap \coNP$ and $\PTIME$ for two-player and one-player games respectively.
\end{restatable}

\subsection{Memory requirements}

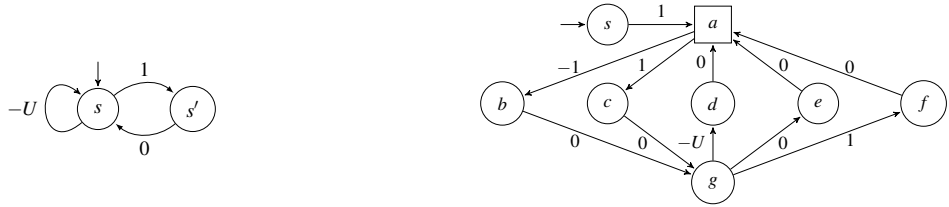
\begin{figure}[tb]
        \centering
\subfloat[$\playerOne$ needs to take $U$ times $(s, s')$ before taking $(s, s)$ once and repeating.]{\label{fig:AELU_memory_p1}\hspace{8mm}\scalebox{0.7}{\begin{tikzpicture}[->,>=stealth',shorten >=1pt,auto,node
    distance=2.5cm,bend angle=45, scale=0.6, font=\normalsize]
    \tikzstyle{p1}=[draw,circle,text centered,minimum size=7mm,text width=4mm]
    \tikzstyle{p2}=[draw,rectangle,text centered,minimum size=7mm,text width=4mm]
    \node[p1]  (0)  at (0, 0) {$s$};
    \node[p1]  (1) at (3, 0) {$s'$};
    \node[]  (empty) at (0, -2.8) {};
    \node[]  (empty) at (-5, 0) {};
    \node[]  (empty) at (5, 0) {};
    
    \coordinate[shift={(0mm,5mm)}] (init) at (0.north);
    \path
    (0) edge [loop left, out=220, in=140,looseness=2, distance=2cm] node [left] {$-U$} (0)
    (init) edge (0);
	\draw[->,>=latex] (0) to[out=40,in=140] node[above] {$1$} (1);
	\draw[->,>=latex] (1) to[out=220,in=320] node[below] {$0$} (0);
      \end{tikzpicture}}\hspace{12mm}}
      \hspace{4mm}
      \subfloat[$\playerTwo$ needs to increase the energy up to $U$ using $(a,c)$ to force $\playerOne$ to take $(g, d)$ then make him lose by taking $(a, b)$.]{\label{fig:AELU_memory_p2}\scalebox{0.7}{\begin{tikzpicture}[->,>=stealth',shorten >=1pt,auto,node
    distance=2.5cm,bend angle=45, scale=1, font=\small]
    \tikzstyle{p1}=[draw,circle,text centered,minimum size=7mm,text width=4mm]
    \tikzstyle{p2}=[draw,rectangle,text centered,minimum size=7mm,text width=4mm]
    \node[p1]  (first)  at (-2, 0) {$s$};
    \node[p2]  (a)  at (0, 0) {$a$};
    \node[p1]  (b) at (-4, -1.5) {$b$};
    \node[p1]  (c) at (-2, -1.5) {$c$};
    \node[p1]  (d) at (0, -1.5) {$d$};
    \node[p1]  (e) at (2, -1.5) {$e$};
    \node[p1]  (f) at (4, -1.5) {$f$};
    \node[p1]  (g) at (0, -3) {$g$};
    \node[]  (empty) at (-6, 0) {};
    \node[]  (empty) at (6, 0) {};
    \coordinate[shift={(-5mm,0mm)}] (init) at (first.west);
    \path
    (first) edge node[above] {$1$} (a)
    (a) edge node[left,xshift=-4mm,yshift=-0.8mm] {$-1$} (b)
    (a) edge node[left,xshift=-1mm] {$1$} (c)
    (d) edge node[left,xshift=0mm] {$0$} (a)
    (e) edge node[right,xshift=1mm] {$0$} (a)
    (f) edge node[right,xshift=4mm,yshift=-0.7mm] {$0$} (a)
    (b) edge node[left,xshift=-4mm,yshift=0.4mm] {$0$} (g)
    (c) edge node[left,xshift=-1mm] {$0$} (g)
    (g) edge node[left,xshift=0mm] {$-U$} (d)
    (g) edge node[right,xshift=1mm] {$0$} (e)
    (g) edge node[right,xshift=4mm,yshift=0.6mm] {$1$} (f)
    (init) edge (first);
      \end{tikzpicture}} }
      \vspace{-1mm}
	\caption{Families of games witnessing the need for pseudo-polynomial-memory strategies for $\EGLU$ (and $\AELU$) objectives. The goal of $\playerOne$ is to keep the energy in $[0,\,U]$ at all times, for $U \in \mathbb{N}$. The left game is won by $\playerOne$ and the right one by $\playerTwo$ but both require memory polynomial in the \textit{value} $U$ to be won.}
	\label{fig:AELU_memory}
\end{figure}

We prove pseudo-polynomial lower and upper bounds on memory for the two players in $\AELU$ games. The upper bound follows from the reduction to a pseudo-polynomial $\AE$ game and the memoryless determinacy of $\AE$ games proved in Thm.~\ref{thm:ae_two_memoryless}.  The lower bound can be witnessed in two families of games asking for strategies using memory polynomial in the energy upper bound $U \in \mathbb{N}$ to be won by $\playerOne$ (Fig.~\ref{fig:AELU_memory_p1}) or $\playerTwo$ (Fig.~\ref{fig:AELU_memory_p2}) respectively. It is interesting to observe that those families already ask for such memory when considering the simpler $\EGLU$ objective.

\begin{restatable}{theorem}{thmAELUmemory}
\label{thm:aelu_memory}
Pseudo-polynomial-memory strategies are both sufficient and necessary to win in $\EGLU$ and $\AELU$ games with arbitrary energy upper bound $U \in \mathbb{N}$, for both players. Polynomial memory suffices when~$U$ is polynomial in the size of the game or encoded in unary.
\end{restatable}
\vspace{-2mm}

\section{Average-Energy with Lower-Bounded Energy}
\label{sec:average_l}

We conclude with the conjunction of an $\AE$ objective with a lower bound (again equal to zero) constraint on the running energy, but no upper bound. This corresponds to an \textit{hypothetical} unbounded energy storage. Hence, its applicability is limited, but it may prove interesting on the theoretical standpoint.

\begin{bproblem}[$\AEL$] 
Given a game~$\Game$, an initial state $\initState$ and a threshold~$t \in \mathbb{Q}$, decide if $\pI$ has a winning strategy $\St_1 \in \strats_{1}$ for objective $\LBound(\initCredit \coloneqq 0)\,\cap\, \AvgEnergyLevel(t)$.
\end{bproblem}

This problem proves to be challenging to solve: we provide partial answers in the following, with a proper algorithm for one-player games but only a correct but incomplete method for two-player games. 

\paragraph{Illustration.} Consider the game in Fig.~\ref{fig:aelu_example}. Recall that for $\AELU$ with $U = 3$, the optimal play is $\play_{3}$, and it requires alternation between all three different simple cycles.
Now consider $\AEL$. One may think that relaxing the objective 
would allow for simpler winning strategies. This is not the case. Some new plays are now acceptable w.r.t.~the energy constraint, such as $\play_{4} = (aabaaba)^{\omega}$, with $\AEsup(\play_{4}) = 11/7$ and $\play_{5} = (aaababa)^{\omega}$, with $\AEsup(\play_{5}) = 18/7$. Yet, the optimal play w.r.t.~the $\AE$ (under the lower-bound energy constraint) is still $\play_{3}$, hence still requires to use all the available cycles, in the appropriate order.

\subsection{One-player games}

We assume that the unique player is $\pI$. Indeed, the opposite case is easy as for $\pII$, the objective is a disjunction and $\pII$ can choose beforehand which sub-objective he will transgress, and do so with a simple memoryless strategy (both $\AEG$ and $\EGL$ games admit memoryless optimal strategies as seen before).
We show how to solve a one-player $\AEL$ problem in \textit{pseudo-polynomial time} by reduction to an $\AELU$ problem for a well-chosen upper bound $U \in \mathbb{N}$ and then application of the algorithm of Sect.~\ref{subsec:AELU_algo}.

\paragraph{The reduction.} Given a game $G = (S_{1}, S_{2} = \emptyset, E, w)$, an initial state $\initState$,  and a threshold $t \in \mathbb{Q}$, we reduce the $\AEL$ problem to an $\AELU$ problem with an upper bound $U \in \mathbb{N}$ which is pseudo-polynomial in the original problem. Precisely, $U \coloneqq t + N^{2} + N^{3}$, with $N = W \cdot (|S|+2)$. The intuition is that if $\playerOne$ can win a one-player $\AEL$ game, he can win it without ever reaching energy levels higher than the chosen bound~$U$, even if he is technically allowed to do so. Essentially, the interest of increasing the energy is making more cycles available (as they become safe to take w.r.t.~the lower bound constraint), but increasing the energy further than necessary is not a good idea as it will negatively impact the average-energy. To prove this reduction, we start from an arbitrary winning path in the $\AEL$ game, and build a witness path that is still winning for the $\AEL$ objective, but
also keeps the energy below $U$ at all times. Our construction exploits a
result of Lafourcade \emph{et~al.} that bounds the value of the counter along a path in a one-counter automaton~\cite{LLT05}. We build upon it to define an appropriate transformation leading to the witness path and derive a sufficiently large upper bound $U \in \mathbb{N}$ for the $\AELU$ problem.

\paragraph{Complexity.} Plugging this bound $U$ in the pseudo-polynomial-time algorithm for $\AELU$ games yields an algorithm for one-player $\AEL$ games that is overall also pseudo-polynomial. We prove that no truly-polynomial-time algorithm can be obtained unless $\PTIME = \NP$ as the one-player $\AEL$ problem is $\NP$-hard. We show it by reduction from the subset-sum problem~\cite{garey_FNY1979}.

\paragraph{Memory requirements.} Recall that for $\playerTwo$, the situation is simpler and memoryless strategies suffice. By the reduction to $\AELU$, we know that pseudo-polynomial memory suffices for $\playerOne$. This bound is tight as witnessed by the family of games already presented in Fig.~\ref{fig:AELU_memory_p1}. To ensure the lower bound on energy, $\playerOne$ has to play edge $(s, s')$ at least $U$ times before taking the $(s, s)$ self-loop. But to minimize the average-energy, edge $(s, s')$ should never be played more than necessary. The optimal strategy is the same as for the $\AELU$ problem: playing $(s, s')$ exactly $U$ times, then $(s, s)$ once, then repeating, forever.

\begin{theorem}
\label{thm:ael_one_memory}
Pseudo-polynomial-memory strategies are both sufficient and necessary to win for $\playerOne$ in one-player $\AEL$ games. Memoryless strategies suffice for $\playerTwo$ in such games.
\end{theorem}

\subsection{Two-player games}

\paragraph{Decidability.} Assume that there exists some $U \in \mathbb{N}$ such that $\playerOne$ has a winning strategy for the $\AELU$ problem with upper bound $U$ and average-energy threshold $t$. Then, this strategy is trivially winning for the $\AEL$ problem as well. This observation leads to an incremental algorithm that is correct (no false positives) but incomplete (it is not guaranteed to stop). In~\cite{BMRLL15a}, we draw the outline of a potential approach to obtain completeness hence decidability.

\begin{lemma}
\label{lem:ael_semi}
There is an algorithm that takes as input an $\AEL$ problem and iteratively solves corresponding $\AELU$ problems for incremental values of $U \in \mathbb{N}$. If a winning strategy is found for some $U \in \mathbb{N}$, then it is also winning for the original $\AEL$ problem. If no strategy is found up to value $U \in \mathbb{N}$, then no strategy of~$\playerOne$ can simultaneously win the $\AEL$ problem and prevent the energy from exceeding~$U$ at all times.
\end{lemma}

While an incomplete algorithm clearly seems limiting from a theoretical standpoint, it~is worth noting that in practice, such approaches are common and often necessary restrictions, even for problems where a complete algorithm is known to exist --- because theoretical bounds granting completeness are too large to be tackled efficiently by software synthesis tools (e.g.,~\cite{CRR14}). In our case, we have already seen that if such a bound exists for the two-player $\AEL$ problem, it needs to be at least exponential in the encoding of problem (cf.~one-player $\AEL$ games). Hence it seems likely that a prohibitive bound would be necessary, rendering the incremental algorithm of Lem.~\ref{lem:ael_semi} more appealing in practice.

\paragraph{Complexity lower bound.} We now prove that the two-player $\AEL$ problem would require at least exponential time to solve. Our proof is by reduction from \textit{countdown games}. A~countdown game
$\mathcal{C}$ is a weighted graph $(\mathcal{V}, \mathcal{E})$, where $\mathcal{V}$ is the finite set of
states, and $\mathcal{E} \subseteq \mathcal{V} \times \mathbb{N} \setminus \{0\} \times \mathcal{V}$ is the edge relation. Configurations are of the form $(v, c)$, $v \in \mathcal{V}$, $c \in \mathbb{N}$. The game starts in an initial configuration $(v_{\text{init}}, c_0)$ and transitions from a configuration $(s, c)$ are performed as follows. First, $\pI$ chooses a duration $d$, $0 < d \leq c$ such that there exists $e = (v, d, v') \in \mathcal{E}$ for some $v' \in \mathcal{V}$. Second, $\pII$ chooses a state $v' \in \mathcal{V}$ such that $e = (v, d, v') \in \mathcal{E}$. Then the game advances to $(v', c-d)$. Terminal configurations are reached whenever no legitimate move is available. If such a configuration is of the form $(v, 0)$, $\pI$ wins the play, otherwise $\pII$ wins. Deciding the winner given an initial configuration $(v_{\text{init}}, c_0)$ is $\EXPTIME$-complete~\cite{JSL08}.

\begin{figure}[tb]
        \centering
\scalebox{0.96}{\begin{tikzpicture}[->,>=stealth',shorten >=1pt,auto,node
    distance=2.5cm,bend angle=45,font=\normalsize,scale=.7,inner sep=.5mm]
    \everymath{\scriptstyle}
    \tikzstyle{p1}=[draw,circle,text centered,minimum size=8mm]
    \tikzstyle{p2}=[draw,rectangle,text centered,minimum size=7mm]
    \node[p1]  (start)  at (-1.5, 1.5) {\small\textsf{start}};
    \node[p1]  (1)  at (0, 0) {$v_{\text{init}}$};
    \node[p2]  (2a) at (3, 1.5) {$(v_{\text{init}}, d_1)$};
    \node[p2]  (2) at (4, 0) {$(v_{\text{init}}, d_2)$};
    \node[p2]  (2b) at (3, -1.5) {$(v_{\text{init}}, d_3)$};
    \node[p1]  (3)  at (8, 0) {$v''$};
    \node[p1]  (3a)  at (7, 1.5) {$v'$};
    \node[p1]  (3b)  at (7, -1.5) {$v'''$};
    \node[p1]  (4) at (-0.5, -2.5) {\small\textsf{stop}};
    \path[use as bounding box] (12,0);
    \coordinate[shift={(-5mm,0mm)}] (init) at (start.west);
    \path
    (init) edge (start)
    (start) edge node[left] {$c_{0}$} (1)
    (1) edge node [left] {$0$} (4)
    (1) edge node [above] {$-d_2$} (2)
    (1) edge node [above,xshift=-1mm] {$-d_1$} (2a)
    (1) edge node [below,xshift=-1mm] {$-d_3$} (2b)
    (2) edge node [above] {$0$} (3)
    (2) edge node [above,xshift=-1mm] {$0$} (3a)
    (2) edge node [below,xshift=-1mm] {$0$} (3b)
    (4) edge [loop left, out=220, in=140,looseness=2, distance=2cm] node [left] {$0$} (4)
    ;
	\draw[->,>=latex] (3) to[out=320, in=355] node [below] {$0$} (4);
	\draw[->,>=latex] (3) to node[above,xshift=-1mm] {$-d_4$} (11,1.5);
	\draw[->,>=latex] (3) to node[above] {$-d_5$} (12,0);
	\draw[->,>=latex] (3) to node[below,xshift=-1mm] {$-d_6$} (11,-1.5);
	\draw[dashed,-,>=latex] (12.3, 0) to (13.3,0);
	\draw[dashed,-,>=latex] (11.3, 1.5) to (12.3,1.5);
	\draw[dashed,-,>=latex] (11.3, -1.5) to (12.3,-1.5);
	\draw[dashed,-,>=latex] (4, 1.5) to (5,1.5);
	\draw[dashed,-,>=latex] (4, -1.5) to (5,-1.5);
	\draw[dashed,-,>=latex] (7.6, 1.5) to (8.6,1.5);
	\draw[dashed,-,>=latex] (7.6, -1.5) to (8.6,-1.5);
\end{tikzpicture}}
\vspace{-1mm}
	\caption{Reduction from a countdown game $\mathcal{C} = (\mathcal{V}, \mathcal{E})$ with initial configuration $(v_{\text{init}}, c_{0})$ to a two-player $\AEL$ problem for average-energy threshold $t \coloneqq 0$.}
\vspace{-4mm}
	\label{fig:countdown_to_ael}
\end{figure}
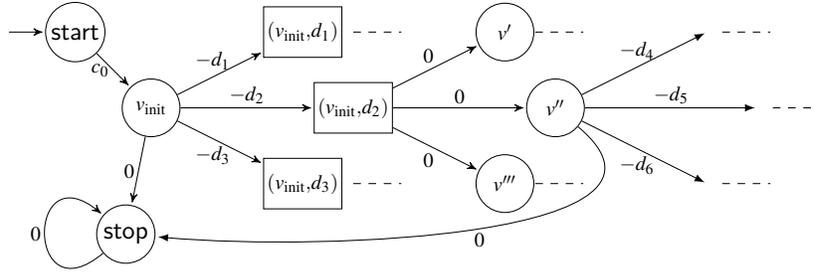

Our reduction is depicted in Fig.~\ref{fig:countdown_to_ael}. The $\EL$ is initialized to $c_0$, then it is decreasing along any play. Consider the $\AEL$ objective for $\AE$ threshold $t \coloneqq 0$. To ensure that the energy always stays non-negative, $\playerOne$ has to switch to \textsf{stop} while the $\EL$ is no less than zero. In addition, to ensure an $\AE$ no more than $t = 0$, $\playerOne$ has to obtain an $\EL$ at most equal to zero before switching to \textsf{stop} (as the $\AE$ will be equal to this $\EL$ thanks to Lem.~\ref{lem:AE_prefix} and the zero self-loop on \textsf{stop}). Hence, $\playerOne$ wins the $\AEL$ objective only if he can ensure a total sum of chosen durations that is \textit{exactly} equal to $c_{0}$, i.e., if he can reach a winning terminal configuration for the countdown game. The converse also holds.

\begin{restatable}{lemma}{lemAELExpHard}
\label{lem:ael_exp_hard}
The $\AEL$ problem is $\EXPTIME$-hard for two-player games.
\end{restatable}

\paragraph{Memory requirements.} We establish that memory is needed for both players.

\begin{restatable}{lemma}{lemAELTwoMemory}
\label{lem:ael_memory}
Pseudo-polynomial-memory strategies are necessary to win for $\playerOne$ in two-player $\AEL$ games. Memory is also required for $\playerTwo$ in such games.
\end{restatable}

\bibliographystyle{new-eptcs}
\bibliography{biblio}

\end{document}